\documentclass[aps,prl,reprint,superscriptaddress]{revtex4-1}

\usepackage{longtable}
\usepackage{morefloats}
\usepackage{color}
\usepackage{graphicx,amsfonts,amsbsy}
\usepackage{amsmath,amsfonts,amsthm,amssymb}
\usepackage{appendix}
\usepackage{makeidx}
\usepackage{url}
\usepackage{verbatim}
\usepackage[bookmarksnumbered,pdfpagelabels=true,plainpages=false,colorlinks=true,linkcolor=blue,citecolor=blue,urlcolor=blue]{hyperref}
\usepackage[rightcaption]{sidecap}
\usepackage{array}
\usepackage{booktabs}
\usepackage{multirow}
\usepackage{tabularx}
\usepackage{cancel,soul,ulem}

\usepackage{mathrsfs}

\newcommand{\bs}[1]{\boldsymbol{#1}}

\begin{document}
\title{Cross-sublattice Spin Pumping and Magnon Level Attraction in van der Waals Antiferromagnets}
\author{Roberto E. Troncoso}
\affiliation{Center for Quantum Spintronics, Department of Physics, Norwegian University of Science and Technology, NO-7491 Trondheim, Norway}

\author{Mike A. Lund}
\affiliation{Department of Engineering Sciences, University of Agder, 4879 Grimstad, Norway}
\affiliation{Center for Quantum Spintronics, Department of Physics, Norwegian University of Science and Technology, NO-7491 Trondheim, Norway}

\author{Arne Brataas}
\affiliation{Center for Quantum Spintronics, Department of Physics, Norwegian University of Science and Technology, NO-7491 Trondheim, Norway}

\author{Akashdeep Kamra}
\affiliation{Center for Quantum Spintronics, Department of Physics, Norwegian University of Science and Technology, NO-7491 Trondheim, Norway}

\begin{abstract}
We theoretically study spin pumping from a layered van der Waals antiferromagnet in its canted ground state into an adjacent normal metal. We find that the resulting dc spin pumping current bears contributions along all spin directions. Our analysis allows for detecting intra- and cross-sublattice spin-mixing conductances via measuring the two in-plane spin current components. We further show that sublattice symmetry-breaking Gilbert damping can be realized via interface engineering and induces a dissipative coupling between the optical and acoustic magnon modes. This realizes magnon level attraction and exceptional points in the system. Furthermore, the dissipative coupling and cross-sublattice spin pumping contrive to produce an unconventional spin current in the out-of-plane direction. Our findings provide a route to extract the spin mixing conductance matrix and uncovers the unique opportunities, such as level attraction, offered by van der Waals antiferromagnet-normal metal hybrids.
\end{abstract}

\maketitle

{\it Introduction}.-- The dawn of magnetic van der Waals (vdW) materials has renewed and invigorated interest in low-dimensional phenomena hosted by solid state systems~\cite{Burch2018,Gibertini2019}. These layered vdWs materials have been found to host various forms of magnetic order~\cite{Huang2017,Deng2018,Huang2018,Song2018,Klein2018,Lado2017} with antiferromagnets (AFs) taking a special place due to their various unique advantages~\cite{jungwirth2016,BaltzRMP2018,Gomonay2014,Gomonay2018}. Among these, a control over interfacial exchange coupling to an adjacent metal and canted or noncollinear ground state offers unprecedented pathways to achieve intriguing physics and applications~\cite{Bender2017,Kamra2017,Troncoso2020,Flebus2019,Hellman2017,Johnsen2020,Rabinovich2019}. The vdW magnets offer an effective control over both - interface and ground state - due to their layered structure and relatively weak interlayer antiferromagnetic exchange~\cite{Huang2017,Deng2018,wang2019}.


Capitalizing on these features, magnon-magnon coupling resulting in level repulsion and hybridization has recently been observed in the vdW AF CrCl$_3$~\cite{MacNeill2019,Kapoor2020,Sklenar2020}. Approaching the challenge from a different direction, similar magnonic hybridization via their mutual coupling~\cite{Kamra2017B,Rezende2019,Yu2020} has been discovered in carefully chosen platforms including compensated ferrimagnets~\cite{Liensberger2019} and synthetic AFs~\cite{Sud2020,Chen2018}. These investigations have, in part, been driven by the desire to control magnonic systems for quantum information applications~\cite{Lachance-Quirion2019} and were preceded by the realization of strong magnon-photon coupling~\cite{Huebl2013,Harder2018B}. The latter, being a relatively mature field, has started to explore dissipative magnon-photon coupling~\cite{Harder2018,Wang2020} resulting in intriguing phenomena that foray into the realm of non-Hermitian physics~\cite{Ashida2020} providing a powerful model platform. While not explored thus far, such phenomena can also result from dissipative magnon-magnon coupling offering various advantages over the magnon-photon platform~\cite{Kamra2020}. Demonstrating these for a vdW AF interfaced with a heavy normal metal (NM) forms a key contribution of this work.


Heterostructures comprising a magnetic insulator interfaced with a thin NM layer have become basic building blocks in an emerging spin-based paradigm for information transport and processing~\cite{Goennenwein2015,Cornelissen2015,Hou2019,Troncoso2020,Lebrun2018,Wimmer2019,Chumak2015,Althammer2018,Nakata2017,Brataas2020}. In such structures, magnonic spin in the magnetic insulator can be interfaced with the electronic spin in NM thereby allowing their integration with conventional electronics. Furthermore, spin generated in NM allows to control, and even negate~\cite{Wimmer2019}, dissipation in the magnetic system via spin transfer torques~\cite{Ralph2008}. Invigorated by recent breakthroughs, especially employing the magnonic spin in AFs~\cite{Lebrun2018,Vaidya160,Li2020,Troncoso2020}, an interface-engineering and control of spin transfer from AF to NM via magnonic spin pumping~\cite{Tserkovnyak2002,Tserkovnyak2005,Cheng2014,Kamra2017,Bender2017} assumes a central role. While coherently driven spin pumping from an AF into NM has recently been observed in its collinear ground state~\cite{Vaidya160,Li2020}, a noncollinear or canted AF should enable unique and novel phenomena emerging from cross-sublattice spin pumping~\cite{Kamra2017,Kamra2018,Liu2017}. Besides providing the much needed understanding of spin transfer across the AF-NM interface, cross-sublattice pumping may also offer a direct probe into the aforementioned dissipative and non-Hermitian magnon-magnon coupling phenomena, as shown in this work.


\begin{figure}[bth]
	\includegraphics[width=85mm]{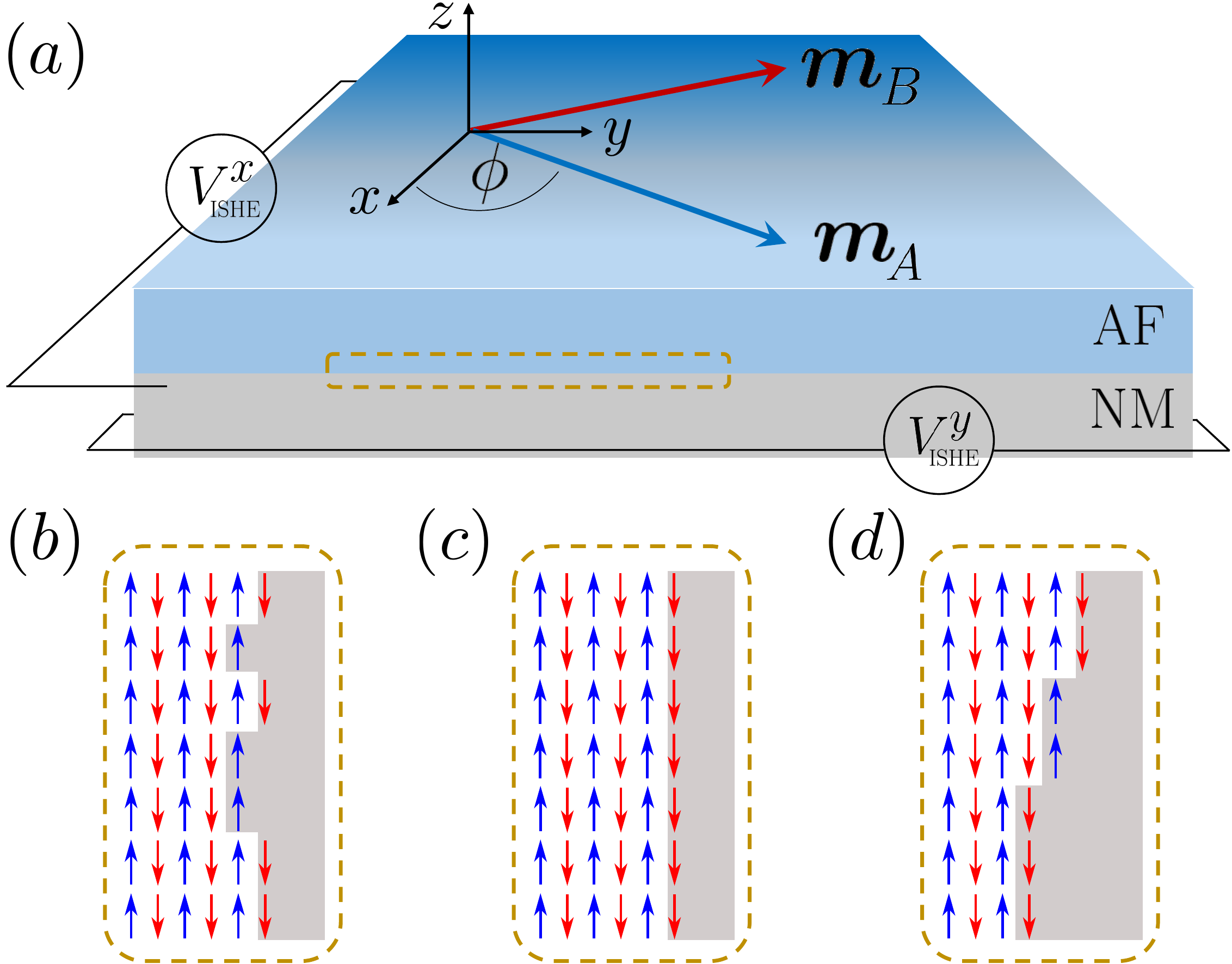}
	\caption{(a) Schematic setup for the measurement of pumped spin currents in the antiferromagnet (AF)-normal metal (NM) structure. The in-plane spin-polarization of the spin current can be detected by the measurements of inverse spin-Hall voltages $V^x_{\text{ISHE}}$ and $V^y_{\text{ISHE}}$. Possible interface microstructures are schematically depicted in (b), (c), and (d). The cross-sublattice spin mixing conductance $g_{AB}$ vanishes [is finite] for the interface depicted in (c) [(b) and (d)].}
	\label{fig:setup}
\end{figure}

In this Letter, we demonstrate heterostructures comprising vdW AFs and a heavy NM to be a unique platform for observing intriguing phenomena emerging from cross-sublattice spin pumping and dissipative magnon-magnon coupling. This niche is enabled by the layered structure of vdW AFs resulting in the possibility of AF-NM interface engineering and canted AF ground states with application of relatively small magnetic fields. We show that in such a non-collinear ground state, the AF pumps spin into the NM along all three directions on excitation via an rf magnetic field. A detection of the two in-plane spin components via inverse spin Hall effect allows to determine the complete spin mixing conductance matrix of the interface. We further find an unconventional out-of-plane spin pumping component that results from a concerted effect of cross-sublattice spin pumping and a dissipative coupling resulting from the sublattice-symmetry breaking at AF-NM interface. Furthermore, the ensuing dissipative coupling is found to result in magnon-magnon level attraction and coalescence observable via typical magnetic resonance experiments. The system thus constitutes a novel and unique platform for investigating this interplay between magnon level repulsion, attraction, and non-Hermitian physics via in-situ damping matrix engineering by, for example, spin transfer torques.


{\it Model}.-- We treat the vdW material as a two-sublattice magnet described by the magnetization fields ${\bs M}_A$ and ${\bs M}_B$ that correspond to the sublattices $A$ and $B$. We consider magnetic free-energy density~\cite{MacNeill2019} $F=\mu_0H_E{\bs M}_A\cdot{\bs M}_B/M_s+{\mu_0}\left(M^2_{Az}+M^2_{Bz}\right)/2-\mu_0{\bs H}\cdot\left({\bs M}_A+{\bs M}_B\right)$, with $M_s$ the saturation magnetization of each sublattice. The inter-sublattice exchange coupling parametrized by $H_E>0$ favors antiferromagnetic order. In addition, an external dc magnetic field ${\bs H}$ is applied in-plane. The second term represents the easy-plane anisotropy. Gilbert damping is accounted for by the viscous Rayleigh dissipation functional~\cite{Gilbert2004,Kamra2018,Yuan2019} via the symmetric matrix $\eta_{\zeta\zeta'}$:
$R[\dot{\bs M}_A,\dot{\bs M}_B]=\sum_{\zeta\zeta'}\int_V d{\bs r}{\eta_{\zeta\zeta'}}\dot{\bs M}_{\zeta}\cdot\dot{\bs M}_{\zeta'}/2$, where $\{\zeta,\zeta'\}=\{A,B\}$. The ensuing magnetization dynamics is described by the coupled Landau-Lifshitz-Gilbert (LLG) equations,
\begin{align}\label{eq:LLG}
\dot{\bs m}_{\zeta}=-\mu_0\gamma{\bs m}_{\zeta}\times {\bs h}^{\text{eff}}_{\zeta}+\alpha_{\zeta\zeta'}{\bs m}_{\zeta}\times\dot{\bs m}_{\zeta'}+{\bs\tau}_{\zeta},
\end{align}
in terms of the unit vectors ${\bs m}_{\zeta} \equiv {\bs M}_{\zeta}/M_s$. The effective fields are given by ${\bs h}^{\text{eff}}_{\zeta} = (1/\mu_0) \times \partial F / \partial {\bs M}_{\zeta} =  {\bs H} -H_{E}\sigma^x_{\zeta\zeta'}{\bs m}_{\zeta'}-M_s\left({\bs m}_{\zeta}\cdot{\hat {\bs z}}\right){\hat {\bs z}}$, with $\sigma^x$ the Pauli matrix and $\gamma>0$ is the gyromagnetic ratio magnitude. The Gilbert damping parameters are defined through $\alpha_{\zeta\zeta'}\equiv \gamma M_s\eta_{\zeta\zeta'}$ where, in particular, $\alpha_{AB}=\alpha_{BA}\equiv\alpha_{od}$. Note that sublattice asymmetry in our model is broken only by the Gilbert damping~\cite{Kamra2018}. It results from the AF-NM interface and spin pumping-mediated losses~\cite{Tserkovnyak2002,Kamra2018}. The magnetization dynamics may be excited by a time-dependent magnetic field ${\bs h}\equiv \mu_0\gamma{\bs h}_{\text{RF}}(t)$ that produces a torque ${\bs{\tau}}_{\zeta}={\bs m}_{\zeta}\times{\bs h}$.

\begin{figure*}[tbh]
  \centering
  \includegraphics[width=\textwidth]{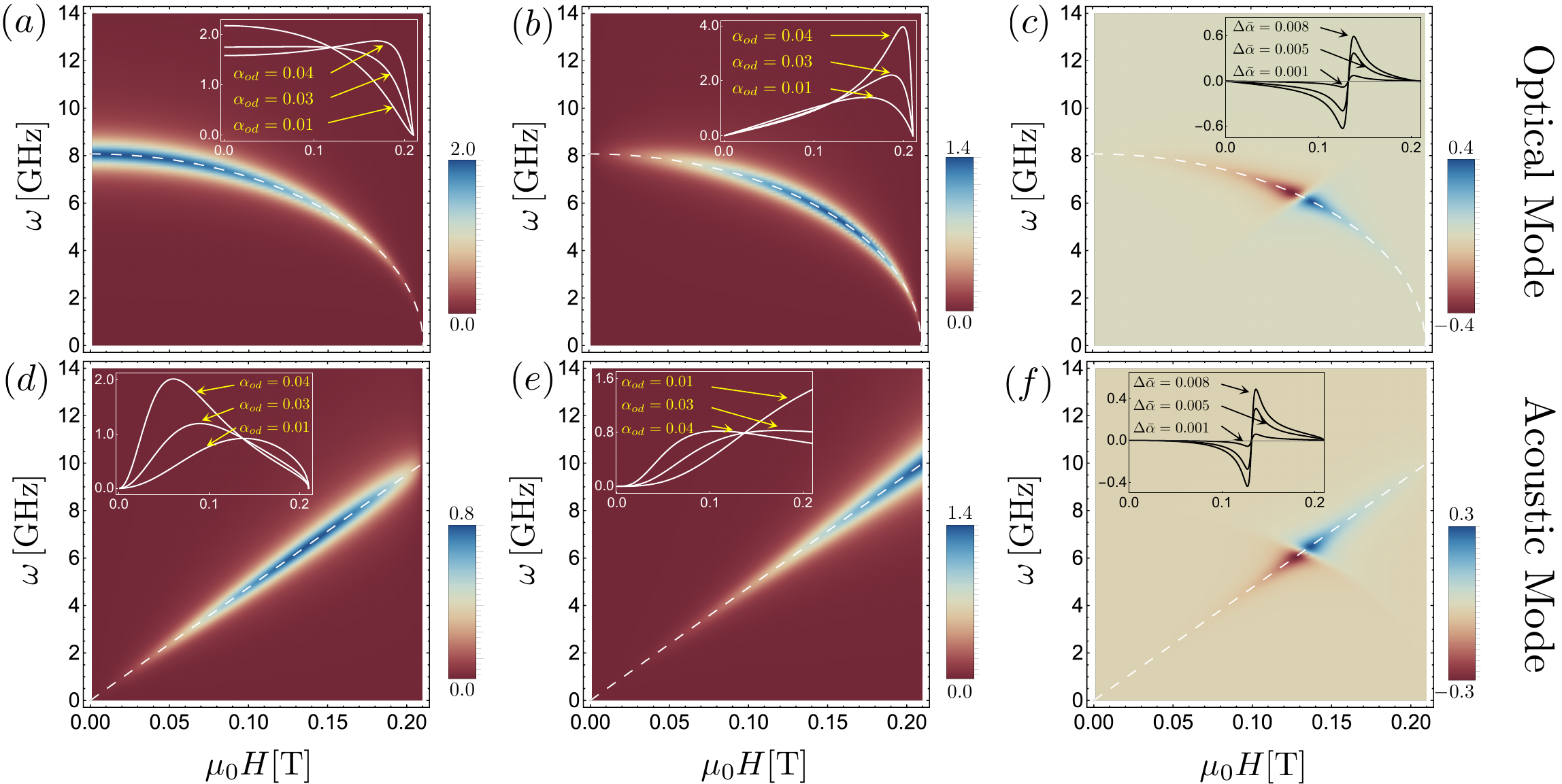}\\
\caption{The frequency and field dependence of the spin-pumping current is represented by the plot of $\mathscr{F}^j_{\pm}(\omega)$, with $j$ the directions of polarization and damping parameters $\bar{\alpha}=0.05$, $\alpha_{od}=0.01$ and $\Delta\bar{\alpha}=0.005$. At panels (a), (b) and (c), we have depicted the spin current with spin-polarization along $x$, $y$- and $z$-direction, respectively, due to coherent excitation of optical modes. Similarly, in panels (d), (e) and (f), we have plotted the polarization components of the spin current when acoustic modes are excited. At the inset of each panel we show $\mathscr{F}^j_{\pm}$, evaluated at the resonant frequencies $\omega=\omega_{\pm}$.  In the inset of panels (a), (b), (d) and (e), the curves correspond to different $\alpha_{od}$, while at panels (c) and (f) $\Delta\bar{\alpha}$ is modified. Other parameters employed at the plots were extracted for the vdW antiferromagnet CrCl$_3$~\cite{MacNeill2019}.}
\label{fig:spinpumping}
\end{figure*}

{\it Magnetization dynamics and magnon modes.--} We now investigate the magnon modes in the material when the two sublattice magnetizations are non-collinear in their equilibrium configuration [Fig.~\ref{fig:setup}(a)]. 
In the presence of an in-plane external magnetic field ${\bs H}=H\hat{\bs y}$, the magnetic ground state becomes ${\bs m}^{\text{eq}}_{\zeta}=\pm\cos\phi\hat{\bs x}+\sin\phi\hat{\bs y}$, where the non-collinearity is captured by the finite angle $\phi$ that satisfies  $\sin\phi=H/2H_E$ [see Fig. \ref{fig:setup}(a)]. The magnetic ground state is invariant under a twofold rotational operation around the $y$ axis, ${\cal C}_{2y}$, in combination with sublattice exchange $A\leftrightarrow B$, i.e. ${\cal C}_{2y}{\bs m}^{\text{eq}}_{A} ={\bs m}^{\text{eq}}_{B}$. Linearizing the LLG equations (\ref{eq:LLG}), considering ${\bs m}_{\zeta}={\bs m}^{\text{eq}}_{\zeta}+\delta{\bs m}_{\zeta}e^{i\omega t}$, the coupled dynamical equations become, 
\begin{subequations}
\begin{align}
i\omega\delta{\bs m}_{+}&\label{eq:opticalmode}={\bs m}^{\text{eq}}_A\times\left({\cal A}_{+}\delta{\bs m}_{+}+i\omega\Delta\bar{\alpha}\delta{\bs m}_{-}\right)+{\bs\tau}_{+},\\
i\omega\delta{\bs m}_{-}&\label{eq:acousticmode}={\bs m}^{\text{eq}}_A\times\left({\cal A}_{-}\delta{\bs m}_{-}+i\omega\Delta\bar{\alpha}\delta{\bs m}_{+}\right)+{\bs\tau}_{-},
\end{align}
\end{subequations}
with the two magnetization dynamics or magnon modes described by the fields $\delta{\bs m}_{\pm}=\delta{\bs m}_A\pm {\cal C}_{2y}\delta{\bs m}_B$, the torques ${\bs\tau}_{\pm}={\bs\tau}_A\pm {\cal C}_{2y}{\bs\tau}_B$ and the operator ${\cal A}_{\pm}=\left(\mu_0\gamma H_E+i\omega\bar{\alpha}\right)\pm\left(\mu_0\gamma H_E+i\omega{\alpha}_{od}\right){\cal C}_{2y}$. Furthermore, we have reformulated the Gilbert damping parameters as $\alpha_{AA}=\bar{\alpha}+\Delta\bar{\alpha}$ and $\alpha_{BB}=\bar{\alpha}-\Delta\bar{\alpha}$. Note that when sublattice symmetry is assumed, i.e., $\alpha_{AA}=\alpha_{BB}$, the Eqs. (\ref{eq:opticalmode}) and (\ref{eq:acousticmode}) become decoupled since $\Delta\bar{\alpha}=0$. In the absence of dissipation, the magnon eigenmodes are captured well by the fields $\delta{\bs m}_{\pm}$, with the eigenfrequencies for the so-called optical and acoustic magnon modes being $\omega_+=\mu_0\gamma \sqrt{2M_sH_E\cos^2\phi}$ and $\omega_-=\mu_0\gamma\sqrt{2H_E\left(M_s+2H_E\right)}\sin\phi$, respectively. The two modes can be excited selectively by a careful choice of the rf-field ${\bs h}$~\cite{MacNeill2019}. In general, the excitation of $(\pm)$-modes one at a time, which demands ${\bs \tau}_{\mp}=0$ for the torque, imposes the conditions ${\bs h}=\pm C_{2y}{\bs h}$, respectively.

{\it Spin pumping}.-- We now investigate spin transport across the AF-NM interface resulting from the excitation of magnetization dynamics by rf magnetic field. The dc spin pumping current injected into the adjacent NM is given by~\cite{Kamra2017} 
\begin{align}\label{eq:sc}
\frac{e}{\hbar}{\bs I}_s=\sum_{\zeta\zeta'\in\{A,B\}}g_{\zeta\zeta'}\langle{\bs m}_{\zeta}\times\dot{\bs m}_{\zeta'}\rangle,   
\end{align}
where $\langle\cdots\rangle$ stands for the time-average over the period of oscillation. The diagonal elements of the matrix $g_{\zeta\zeta'}$ describe the intra-sublattice spin mixing conductance. An asymmetric interfacial coupling~\cite{Bender2017,Kamra2017,Troncoso2020,Flebus2019}, resulting in $g_{AA}\neq g_{BB}$, occurs when the two magnetic sublattices are incommensurately exposed to the NM (see Fig.~\ref{fig:setup}(c) for an example in which $g_{AA} = 0$). The off-diagonal cross-sublattice conductance satisfy $g_{AB}=g_{BA}$ and are nonzero when both the sublattices are (partly) exposed to NM [Fig.~\ref{fig:setup}(b) and (d)]. Such interfaces can be achieved with layered magnets~\cite{Yin2018,Huang2018} (Fig.~\ref{fig:setup}), but are not possible with synthetic AFs~\cite{Sud2020}. Our goal here is to establish the experimentally detectable spin pumping current as a direct probe of the $2 \times 2$ spin mixing conductance matrix [Eq. (\ref{eq:sc})]. In particular, we are interested in establishing unique signatures of the cross-sublattice conductances $g_{AB} = g_{BA}$ that elude a direct experimental observation thus far.


In typical experimental setups (Fig. \ref{fig:setup}), the spin pumping current is detected via inverse spin-Hall effect (ISHE)~\cite{Sinova2015,Ando2011,Valenzuela2006,Saitoh2006}. In metals with strong spin-orbit coupling, a nonequilibrium spin current ${\bs j}_s$ induces a transverse charge current ${\bs j}_c=\theta_{\text{SH}}(2e/\hbar){\bs j}_s\times\vec{\sigma}$, where $\vec{\sigma}$ denotes the spin polarization direction and $\theta_{\text{SH}}$ is the spin Hall angle~\cite{Sinova2015}. Under open circuit conditions, the generated charge current is countered by an induced inverse spin Hall voltage $V_{\text{ISHE}}$ proportional to the charge, and thus spin, current. The voltage thus generated is proportional to the spin current injected into the NM [Eq.~\eqref{eq:sc}] and is directly detected in experiments~\cite{Mosendz2010}.

In order to relate spin mixing conductance matrix elements to experimental observables, we evaluate the spin pumping current Eq. (\ref{eq:sc}) in the limit $\alpha_{AA}\approx\alpha_{BB}$, i.e., $\Delta\bar{\alpha}$ small. In this perturbative regime, Eqs. (\ref{eq:opticalmode}) and (\ref{eq:acousticmode}) decouple and the system eigenmodes are the optical and acoustic magnon modes. The corresponding spin pumping currents are evaluated to be
\begin{align}\label{eq:scpm}
{\bf I}^{\pm}_z\nonumber=&\left(g_{AA}-g_{BB}\right)\mathscr{F}^{x}_{\pm}(\omega)\hat{\bs x}+g_{AB}\mathscr{F}^{z}_{\pm}(\omega)\hat{\bs z}\\
&\qquad\qquad+\left(g_{AA}+g_{BB}\pm 2g_{AB}\right)\mathscr{F}^{y}_{\pm}(\omega)\hat{\bs y}.
\end{align}
The $\pm$ index labels optical and acoustic modes, which are driven by the rf magnetic field with frequency $\omega$ and amplitude $h_{+,\phi}=2h_y\cos\phi$ and $h_{-,\phi}= 2h_x\sin\phi$, respectively. The expression in Eq. (\ref{eq:scpm}) represents a pure spin current that flows across the AF-NM interface (along $z$ axis). Its various components pertaining to directions in the spin space are proportional to the functions $\mathscr{F}^{j}_{\pm}(\omega)$, as detailed in the Supplemental Material  in \cite{SM} (see Eqs. (\ref{eqFx})-(\ref{eqFz})), which have been obtained to the first order in $\Delta\bar{\alpha}$. The in-plane components $\mathscr{F}^{x,y}_{\pm}$, satisfying $\mathscr{F}^{y}_{\pm}(\omega)=\tan\phi\mathscr{F}^{x}_{\pm}(\omega)$, are independent of $\Delta\bar{\alpha}$. However, the out-of-plane component of the spin current $\propto \mathscr{F}^{z}_{\pm}(\omega)$ scales linearly with $\Delta\bar{\alpha}$. The resulting field- and frequency-dependence of the spin current ${\bf I}^{\pm}_z$ components are plotted in Fig.~\ref{fig:spinpumping} employing system parameters relevant for the vdW AF CrCl$_3$~\cite{MacNeill2019}. The various spin current components are displayed in panels (a), (b) and (c) of Fig. \ref{fig:spinpumping} for the optical mode, and in panels (d), (e), and (f) for the acoustic mode.

In contrast with the case of spin pumping via collinear magnets, in which a dc spin current polarized along the equilibrium order is generated~\cite{Tserkovnyak2002,Cheng2014,Mosendz2010,Saitoh2006,Vaidya160,Li2020}, the canted AF under consideration pumps spin with components along all three directions [Eq.~\eqref{eq:scpm}]. As detailed in the Supplemental Material~\cite{SM}, $\mathscr{F}^{y,z}_{\pm} \propto \sin \phi$ implying $y$ and $z$ components of the spin pumping current vanish for $\phi = 0$. Our result thus reduces to the existing understanding of collinear AFs~\cite{Kamra2017,Troncoso2020}. These additional components of the spin pumping current for the canted AF together with the existence of two independent magnon modes constitute some of the unique features and opportunities offered by this system. For example, detection of the ISHE voltage in two orthogonal directions [as depicted in Fig.~\ref{fig:setup}(a)] enables determination of both in-plane spin current components. By detecting these while exciting the two magnon modes one at a time, we may determine the full spin mixing conductance matrix with the cross-sublattice term given by
\begin{align}\label{eq:gAB}
{g_{AB}}=\frac{g_{AA}-g_{BB}}{4\tan\phi}\left(\frac{{I}^{+}_{s,y}}{{I}^{+}_{s,x}}-\frac{{I}^{-}_{s,y}}{{I}^{-}_{s,x}}\right),
\end{align}
which assumes the condition $g_{AA}\neq g_{BB}$. This accomplishes a key goal and constitutes a main result of this paper. 

Furthermore, existence of the spin current $z$ component [Eq.~\eqref{eq:scpm}] that we find is unconventional and counterintuitive as the magnon spin is expected to lie in the same plane as the equilibrium sublattice magnetizations~\cite{Kamra2017B,Rezende2019}. However, this component is nonzero only when $\Delta\bar{\alpha} \neq 0$, $g_{AA}\neq g_{BB}$, $g_{AB} \neq 0$, and $\phi \neq 0$ implying that it results from a complex interplay of the sublattice-symmetry breaking dissipative coupling and a cross-sublattice interference effect. Such physics, especially dissipative coupling~\cite{Harder2018,Wang2020}, appears to go beyond the magnon picture considered thus far and constitutes another key result of our work. While an out-of-plane spin component has not been measured in typical spin pumping experiments~\cite{Saitoh2006,Sinova2015,Mosendz2010}, the recent thermal drag-mediated detection of such an out-of-plane spin component~\cite{Avci2020} provides one possible method for its direct observation.

\begin{figure}[tbh]
	\includegraphics[width=\linewidth,clip=]{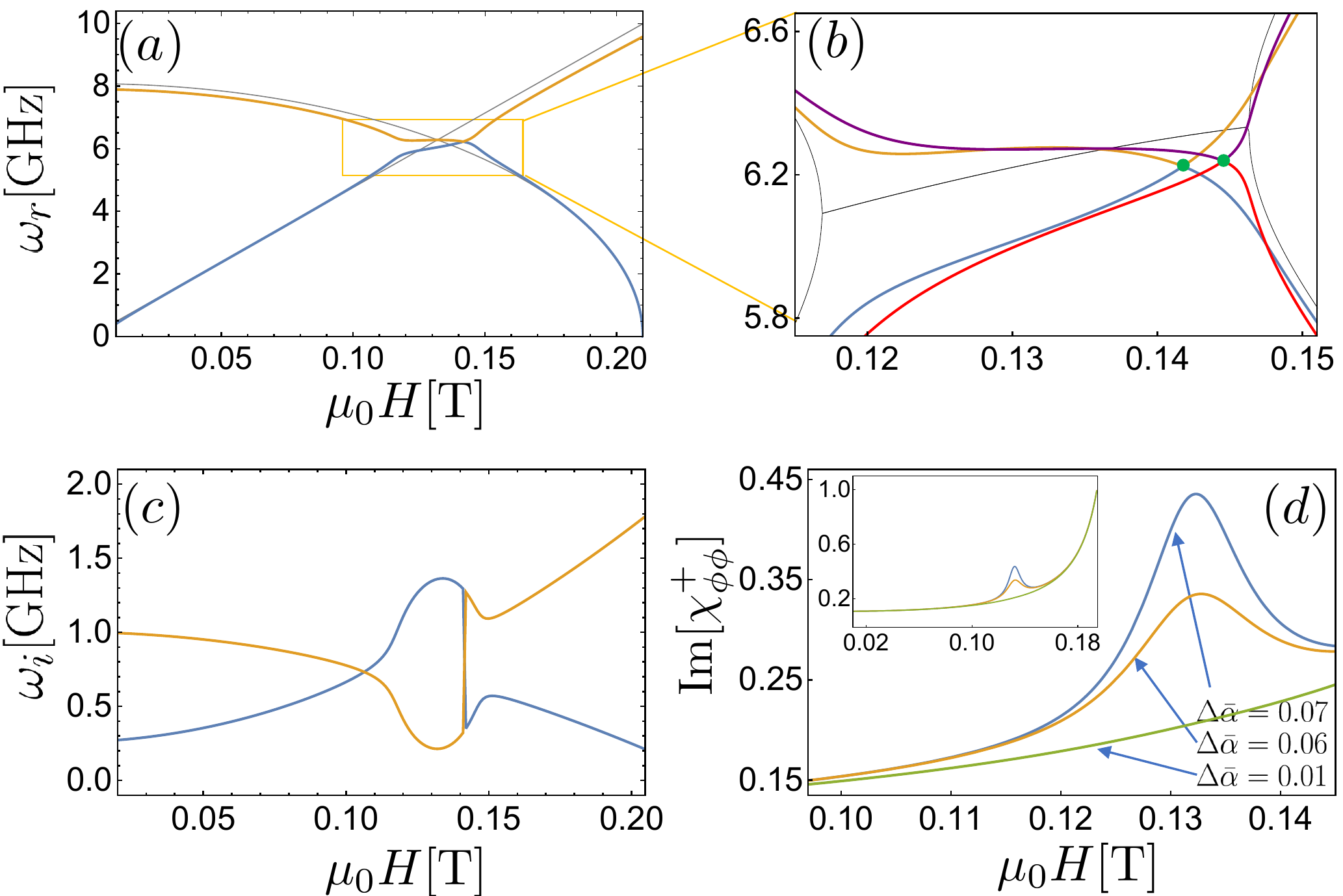}
	\caption{Eigenfrequencies of the coupled magnon modes are plotted as a function of the external dc magnetic field. The real ($\omega_r$) and imaginary ($\omega_i$) parts of the frequencies are displayed in panels (a) and (c) respectively, for $\bar{\alpha}=0.1$ and $\alpha_{od}=\Delta\bar{\alpha}=0.07$. In panel (b), we zoom-in on the level crossing of panel (a). The additional curves shown in purple and red correspond to the same $\bar{\alpha}$ and $\Delta\bar{\alpha}$ as (a), but with $\alpha_{od}=0$. The level attraction is depicted in gray when $\Delta\bar{\alpha}=0.07$ and $\bar{\alpha}=\alpha_{od}=0$. (d) Imaginary part of the magnetic susceptibility $\text{Im}[\chi^{+}_{\phi\phi}]$ {\it vs.} applied magnetic field for various values of $\Delta\bar{\alpha}$. The susceptibility corresponds to the optical mode and has been evaluated at $\omega=\omega_+$.}
	\label{fig:hybridization}
\end{figure}

{\it Magnon level attraction}.-- We now discuss the magnon eigenmodes which become dissipatively coupled [see Eqs.~\eqref{eq:opticalmode} and \eqref{eq:acousticmode}] due to the sublattice-symmetry breaking Gilbert damping~\cite{Kamra2018}, i.e. nonzero $\Delta\bar{\alpha}$. While a ``reactive'' coupling between magnon modes has been observed in various systems~\cite{Chen2018,Liensberger2019,Sud2020,MacNeill2019,Kapoor2020}, dissipative coupling remains less explored and invokes non-Hermitian physics~\cite{Ashida2020,Flebus2020}. Solving Eqs.~\eqref{eq:opticalmode} and \eqref{eq:acousticmode} without an external rf drive, we obtain the complex eigenmode frequencies $\omega_{\pm} = \omega_{r \pm} + i \omega_{i\pm}$. $\omega_{r}$ and $\omega_{i}$, respectively, capture the energy and inverse lifetime of the magnon modes and have been plotted against the external dc magnetic field in Fig.~\ref{fig:hybridization} (a)-(c). Due to the dissipative nature of the coupling, a magnon-magnon level attraction is observed. The grey curve in Fig. \ref{fig:hybridization} (b), corresponding to $\bar{\alpha}=\alpha_{od}=0$ and $\Delta\bar{\alpha}\neq 0$, depicts a perfect level coalescence or mode synchronization~\cite{Harder2018,Wang2020}. Such values for Gilbert damping matrix require a dc spin transfer torque drive in the NM. An undriven system however imposes constraints $\alpha_{AA}, \alpha_{BB} > 0$, i.e. $\bar{\alpha}>\Delta\bar{\alpha}$ and $\alpha_{od}\leq \sqrt{\alpha_{AA}\alpha_{BB}}$~\cite{Kamra2018}. In this scenario, we find a complex interplay of repulsion and attraction between the two modes. The resulting eigenfrequencies split slightly (see Fig. \ref{fig:hybridization}(a) when $\bar{\alpha}=0.1$, $\alpha_{od}=0.07$ and $\Delta\bar{\alpha}=0.07$) while coalescing at a specific point, the so-called exceptional point~\cite{Tserkovnyak2020}. Furthermore, Fig. \ref{fig:hybridization}(d) depicts the imaginary part of dynamic susceptibility, which is directly accessible in experiments~\cite{Liensberger2019,MacNeill2019,Berger2018}. A peak in this susceptibility provides an additional experimental signature of the level attraction when $\Delta \bar{\alpha}$ is sufficiently large. Thus, vdW AFs under consideration constitute a rich platform for realizing non-Hermitian physics and magnon-magnon level attraction via AF-NM interface engineering and spin transfer torques exerted on AF by the NM.

{\it Summary}.-- We have theoretically uncovered unique and intriguing cross-sublattice spin pumping and magnon-magnon level attraction effects in a model canted antiferromagnet. By providing guidance to experiments in extracting the interfacial spin mixing conductance matrix and key signatures of level attraction, we hope to establish van der Waals antiferromagnets interfaced with a heavy metal layer as a fertile and convenient platform for realizing and investigating unconventional non-Hermitian physics.

Note added: During the manuscript preparation, we noticed a recent related preprint~\cite{Lu2020} that studies level-repulsion and hybridization of magnonic modes in bulk symmetry-breaking synthetic antiferromagnets. It however does not discuss spin pumping or the dissipative magnon level attraction - the two key novelties of our work.
\begin{acknowledgments}
This work was supported by the European Union’s Horizon 2020 Research and Innovation Programme under Grant No. DLV-737038 “TRANSPIRE,” and the Research Council of Norway through its Centres of Excellence funding scheme, Project No. 262633, ``QuSpin''.
\end{acknowledgments}

\bibliography{CSSPvdWAF.bib}

\clearpage
\onecolumngrid
\section{Supplemental Material}
In this Supplemental Material, we explicitly show the calculation of two-sublattice spin pumping in AF-NM structures. The result is applied when the magnetization dynamics is coherently driven by a linearly polarized ac magnetic field. Basic details  of magnon-magnon hybridization are also provided.

\subsection{Magnetization dynamics and magnon modes}
Magnetic fluctuations of the two-sublattice magnet results from the linearization of LLG equations. Representing the fluctuations by ${\bs m}_{\zeta}={\bs m}^{\text{eq}}_{\zeta}+\delta{\bs m}_{\zeta}e^{i\omega t}$, with $\zeta=\{A,B\}$, the resulting coupled dynamical equations becomes, 
\begin{subequations}
	\begin{align}
		i\omega\delta{\bs m}_{+}&\label{eq:opticalM}={\bs m}^{\text{eq}}_A\times\left[\left(\mu_0\gamma H_E+i\omega\bar{\alpha}\right)\delta{\bs m}_{+}+\left(\mu_0\gamma H_E+i\omega{\alpha}_{od}\right){\cal C}_{2y}\delta{\bs m}_{+}\right]+i\omega\Delta\bar{\alpha}{\bs m}^{\text{eq}}_A\times\delta{\bs m}_{-}+{\bs\tau}_{+},\\
		i\omega\delta{\bs m}_{-}&\label{eq:acousticM}={\bs m}^{\text{eq}}_A\times\left[\left(\mu_0\gamma H_E+i\omega\bar{\alpha}\right)\delta{\bs m}_{-}-\left(\mu_0\gamma H_E+i\omega{\alpha}_{od}\right){\cal C}_{2y}\delta{\bs m}_{-}\right]+i\omega\Delta\bar{\alpha}{\bs m}^{\text{eq}}_A\times\delta{\bs m}_{+}+{\bs\tau}_{-},
	\end{align}
\end{subequations}
with the torques ${\bs\tau}_{\pm}={\bs m}_A\times{\bs h}_{\pm}$. The Gilbert damping terms are represented by $\alpha_{AB}=\alpha_{BA}=\alpha_{od}$, $\alpha_{AA}=\bar{\alpha}+\Delta\bar{\alpha}$ and $\alpha_{BB}=\bar{\alpha}-\Delta\bar{\alpha}$. Note that when sublattice symmetry is restored, i.e., $\alpha_{AA}=\alpha_{BB}$, the modes $ \delta{\bs m}_{+}$ and $\delta{\bs m}_{-}$  become decoupled and characterize, optical and acoustic modes respectively. In a compact form the Eqs. (\ref{eq:opticalM}) and (\ref{eq:acousticM}) read,
\begin{align}\label{eq:deltah}
	\left(\begin{array}{c}
		{\bs h}_{+} \\
		{\bs h}_{-}
	\end{array}\right)=\underbrace{\left(
		\begin{array}{c|c}
			\mathbb{M}_+ & \mathbb{T}\\
			\hline 
			\mathbb{T}^* & \mathbb{M}_-
		\end{array}\right)}_{\mathbb{M}}
	\left(\begin{array}{c}
		\delta{\bs m}_{+} \\
		\delta{\bs m}_{-}
	\end{array}\right)
\end{align} 
with the ac fields ${\bs h}_{+}=(h_{+,\phi},0)^T$ and ${\bs h}_{-}=(h_{-,\phi},h_{-,\theta})^T$, where $h_{+,\phi}=2h_y\cos\phi$, $h_{-,\phi}= 2h_x\sin\phi$ and $h_{-,\theta}= 2h_z$. The matrices $\mathbb{M}_{\pm}$ and $\mathbb{T}$ are defined as
\begin{align}
		\mathbb{T}&=\left(\begin{array}{cc}
		-i\omega\Delta\bar{\alpha} & 0\\
		0 & i\omega\Delta\bar{\alpha}
	\end{array} \right),\\
	\mathbb{M}_+&=
	\left(\begin{array}{cc}
		-A-i\omega\left(\bar{\alpha}+\alpha_{od}\cos 2\phi\right) &i\omega \\
		i\omega & B+i\omega\left(\bar{\alpha}-\alpha_{od}\right)
	\end{array} \right),\\
	\mathbb{M}_-&=\left(\begin{array}{cc}
		C+i\omega\left(\bar{\alpha}-\alpha_{od}\cos 2\phi\right) & -i\omega\\
		-i\omega & -D-i\omega\left(\bar{\alpha}+\alpha_{od}\right)
	\end{array} \right),
\end{align}
where we introduced the following constants $A=2\mu_0\gamma H_E\cos^2\phi$ and $B=\mu_0\gamma M_s$, $C=2\mu_0\gamma H_E\sin^2\phi$ and $D=\mu_0\gamma \left(M_s+2H_E\right)$.  It is worth noting that $\det\left[\mathbb{M}\right]=\det\left[\mathbb{M}_+\right]\det\left[\mathbb{M}_--\mathbb{T}\mathbb{M}^{-1}_+\mathbb{T}\right]\approx \det\left[\mathbb{M}_+\right]\det\left[\mathbb{M}_-\right]$ in the limit of small $\Delta\bar{\alpha}$. Beyond this approximation, the optical and acoustic magnonic modes are no longer decoupled. Instead, these modes hybridize with $\Delta\bar{\alpha}$ being the dissipative coupling that generates level attraction. The magnon-magnon hybrid eigenfrequencies were numerically calculated and depicted in Fig. \ref{fig:hybridization}.

Assuming that $\Delta\bar{\alpha}$ is small, we use the standard formula for  the inverse of a block matrix to obtain the fields $\delta{\bs m}_{\pm}$,
\begin{align}
	\delta{\bs m}_{+}&\label{eq:deltamplus}=\mathbb{M}^{-1}_+{\bs h}_+  - \mathbb{M}^{-1}_+\mathbb{T} \mathbb{M}^{-1}_-{\bs h}_-,\\
	\delta{\bs m}_{-}&\label{eq:deltamminus}=\mathbb{M}^{-1}_-{\bs h}_-  + \mathbb{M}^{-1}_-\mathbb{T} \mathbb{M}^{-1}_+{\bs h}_+.
\end{align} 
The first term at the right-hand side is  the zero order correction in $\Delta\bar{\alpha}$, where $\mathbb{M}^{-1}_{\pm}$ corresponds to the symmetric dynamic susceptibility matrix.  The components are defined as $\left[\mathbb{M}^{-1}_{\pm}\right]_{11}=\chi^{\pm}_{\phi\phi}$, $\left[\mathbb{M}^{-1}_{\pm}\right]_{22}=\chi^{\pm}_{\theta\theta}$ and $\left[\mathbb{M}^{-1}_{\pm}\right]_{12}=i\chi^{\pm}_{\phi\theta}$, where
\begin{align}
	\chi^+_{\phi\phi}&\label{eq:suscepp1}=\frac{B+i\left(\bar{\alpha}-\alpha_{od}\right)\omega}{\omega^2-\left(B+i\omega(\bar{\alpha}-\alpha_{od})\right)\left(A+i\omega(\bar{\alpha}+\alpha_{od}\cos 2\phi)\right)},\\
	\chi^+_{\phi\theta}&\label{eq:suscepp2}=\frac{-\omega}{\omega^2-\left(B+i\omega(\bar{\alpha}-\alpha_{od})\right)\left(A+i\omega(\bar{\alpha}+\alpha_{od}\cos 2\phi)\right)},
\end{align}
and
\begin{align}
	\chi^-_{\phi\phi}&\label{eq:suscepm1}=\frac{-D-i\omega(\bar{\alpha}+\alpha_{od})}{\omega^2-\left(D+i\omega(\bar{\alpha}+\alpha_{od})\right)\left(C+i\omega(\bar{\alpha}-\alpha_{od}\cos 2\phi)\right)},\\
	\chi^-_{\theta\theta}&\label{eq:suscepm2}=\frac{C+i\omega(\bar{\alpha}-\alpha_{od}\cos 2\phi)}{\omega^2-\left(D+i\omega(\bar{\alpha}+\alpha_{od})\right)\left(C+i\omega(\bar{\alpha}-\alpha_{od}\cos 2\phi)\right)},\\
	\chi^-_{\phi\theta}&\label{eq:suscepm3}=\frac{\omega}{\omega^2-\left(D+i\omega(\bar{\alpha}+\alpha_{od})\right)\left(C+i\omega(\bar{\alpha}-\alpha_{od}\cos 2\phi)\right)}.
\end{align}

The second contribution in Eqs. (\ref{eq:deltamplus}) and (\ref{eq:deltamminus}) is linear in the damping $\Delta\bar{\alpha}$, with the $2\times 2$ matrices,
\begin{align}
	\mathbb{M}^{-1}_+\mathbb{T} \mathbb{M}^{-1}_-&=\frac{\omega\Delta\alpha}{\det\left[\mathbb{M}_+\right]\det\left[\mathbb{M}_-\right]}\left(\begin{array}{cc}
		i\left(BD+\omega^2\right)& \left(B+C\right)\omega \\
		\left(A+D\right)\omega& -i \left(AC+\omega^2\right)
	\end{array}\right),\\
	\mathbb{M}^{-1}_-\mathbb{T} \mathbb{M}^{-1}_+&=\frac{\omega\Delta\alpha}{\det\left[\mathbb{M}_+\right]\det\left[\mathbb{M}_-\right]}\left(\begin{array}{cc}
		i\left(BD+\omega^2\right)& \left(A+D\right)\omega \\
		\left(B+C\right)\omega& -i \left(AC+\omega^2\right)
	\end{array}\right).
\end{align}

\subsection{Two-sublattice Spin Pumping}

In this section we evaluate the spin pumping in the two-sublattice magnet. Final expressions for the injected spin currents are found in terms of the dynamical magnetic susceptibility. To start with, let us consider the spin pumping current into the normal metal given by Eq. (\ref{eq:sc}).  The fluctuations are represented by ${\bs m}_{\zeta}={\bs m}^{\text{eq}}_{\zeta}+\delta{\bs{\cal M}}_{\zeta}$, with the real-valued fields defined as $\delta{\bs{\cal M}}_{A}=\delta{\cal M}_{A,\theta}\hat{\bs z}+\delta{\cal M}_{A,\phi}\left(\hat{\bs z}\times{\bs m}^{\text{eq}}_A\right)$ and $\delta{\bs{\cal M}}_{B}=\delta{\cal M}_{B,\theta}\hat{\bs z}+\delta{\cal M}_{B,\phi}\left(\hat{\bs z}\times{\bs m}^{\text{eq}}_B\right)$. We find that the spin pumping current ${\bf I}_s$ becomes
\begin{align}\label{eq:spgen}
	\frac{4\pi}{\hbar}{\bf I}_s=g_{AA}\langle\delta{\bs{\cal M}}_A\times\delta\dot{\bs{\cal M}}_A\rangle+g_{AB}\left(\langle\delta{\bs{\cal M}}_A\times\delta\dot{\bs{\cal M}}_B\rangle+\langle\delta{\bs{\cal M}}_B\times\delta\dot{\bs{\cal M}}_A\rangle\right)+g_{BB}\langle\delta{\bs{\cal M}}_B\times\delta\dot{\bs{\cal M}}_B\rangle.
\end{align}
Note that since a time-average is involved in the evaluation of previous equation, linear terms in the fluctuations do not contribute. In order to relate each term in Eq. (\ref{eq:spgen}) with the dynamical susceptibility, Eqs. (\ref{eq:suscepp1})-(\ref{eq:suscepm3}), we represent $\delta{\bs{\cal M}}_{\zeta}$ by complex-valued fields as $\delta{{\cal M}}_{\zeta,\mu}=\text{Re}\left[\delta{m}_{\zeta,\mu}e^{i\omega t}\right]$, where $\mu=\{\theta,\phi\}$. Next, we write the fields $\delta{m}_{\zeta,\mu}$ in the eigenbasis by the following relations $\delta m_{A/B,\theta}=\frac{1}{2}\left(\delta m_{\pm,\theta}\pm\delta m_{\mp,\theta}\right)$ and $\delta m_{A/B,\phi}=\frac{1}{2}\left(\delta m_{\pm,\phi}\pm\delta m_{\mp,\phi}\right)$. Therefore, we obtain 
\begin{align}
&\langle\delta{\bs{\cal M}}_A\times\delta\dot{\bs{\cal M}}_A\rangle=\frac{\omega}{4}\left(\text{Im}\left[\langle\delta m_{+,\phi}\delta m^*_{+,\theta}\rangle\right]+\text{Im}\left[\langle\delta m_{-,\phi}\delta m^*_{-,\theta}\rangle\right]+\text{Im}\left[\langle\delta m_{-,\phi}\delta m^*_{+,\theta}\rangle\right]\label{eq:spAA}+\text{Im}\left[\langle\delta m_{+,\phi}\delta m^*_{-,\theta}\rangle\right]\right){\bs m}^{\text{eq}}_A,\\
&\label{eq:spBB}\langle\delta{\bs{\cal M}}_B\times\delta\dot{\bs{\cal M}}_B\rangle=\frac{\omega}{4}\left(\text{Im}\left[\langle\delta m_{-,\phi}\delta m^*_{-,\theta}\rangle\right]+\text{Im}\left[\langle\delta m_{+,\phi}\delta m^*_{+,\theta}\rangle\right]-\text{Im}\left[\langle\delta m_{-,\phi}\delta m^*_{+,\theta}\rangle\right]-\text{Im}\left[\langle\delta m_{+,\phi}\delta m^*_{-,\theta}\rangle\right]\right){\bs m}^{\text{eq}}_B,\\
&\nonumber\langle\delta{\bs{\cal M}}_A\times\delta\dot{\bs{\cal M}}_B\rangle+\langle\delta{\bs{\cal M}}_B\times\delta\dot{\bs{\cal M}}_A\rangle={\frac{\omega}{2}}\cos\phi\left(\text{Im}\left[\langle\delta m_{+,\phi}\delta m^*_{-,\theta}\rangle\right]
	-\text{Im}\left[\langle\delta m_{-,\phi}\delta m^*_{+,\theta}\rangle\right]\right)\hat{\bs x}\\
	&\qquad\qquad\qquad\qquad+{\frac{\omega}{2}}\sin\phi\left(\text{Im}\left[\langle\delta m_{-,\phi}\delta m^*_{-,\theta}\rangle\right]
	-\text{Im}\left[\langle\delta m_{+,\phi}\delta m^*_{+,\theta}\rangle\right]\right)\hat{\bs y}\label{eq:spAB}+{\frac{\omega}{2}}\sin(2\phi)\text{Im}\left[\langle\delta m_{+,\phi}\delta m^*_{-,\phi}\rangle\right]\hat{\bs z}.
\end{align}

Below, we present examples of spin pumping current, evaluating Eqs. (\ref{eq:spAA}-\ref{eq:spAB}), due to the coherent magnonic excitation by linearly polarized external ac fields.

\subsubsection{Exciting the Optical and Acoustic Modes}\label{app:opticalacousticmodes}
According to the selection rules described in the main text, excitation of the optical mode requires a field that satisfy ${\bs h}_{-}={\bs 0}$ and $h_{+,\phi}\neq 0$. On the other hand, excitation of the acoustic mode requires ${\bs h}_{+}={0}$ and ${\bs h}_{-}\neq 0$. From Eqs. (\ref{eq:deltamplus}) and (\ref{eq:deltamminus}), we find the fields $\delta{\bs m}_{\pm}$ when the ac magnetic field is ${\bs h}_+=h_y\hat{\bs y}$ (optical mode) and ${\bs h}_-=h_x\hat{\bs x}$ (acoustic mode). To determine the spin current,  we evaluate each term given by the Eqs. (\ref{eq:spAB})-(\ref{eq:spBB}). Up to linear order in the Gilbert damping difference $\Delta\bar{\alpha}$ , we find
\begin{align}\label{eq:scresult}
	\frac{e}{\hbar}{\bf I}^{\pm}_s=
	\frac{\omega}{4}\text{Im}\left[\langle\delta m_{\pm,\phi}\delta m^*_{\pm,\theta}\rangle\right]&\nonumber\left[\left(g_{AA}-g_{BB}\right)\cos\phi\hat{\bs x}+\left(\left(g_{AA}+g_{BB}\right)\mp 2g_{AB}\right)\sin\phi\hat{\bs y}\right]\\
	&\qquad\qquad\qquad\qquad\qquad+\frac{\omega}{2} \text{Im}\left[\langle\delta m_{+,\phi}\delta m^*_{-,\phi}\rangle\right]_{\pm}g_{AB}\sin(2\phi)\hat{\bs z}.
\end{align}
The functions $\mathscr{F}^{j}_{\pm}(\omega)$ introduced in the main text, are thus defined as
\begin{align}
	\mathscr{F}^{x}_{\pm}(\omega)&\label{eqFx}=	\frac{\omega}{4}\text{Im}\left[\langle\delta m_{\pm,\phi}\delta m^*_{\pm,\theta}\rangle\right]\cos\phi,\\
	\mathscr{F}^{y}_{\pm}(\omega)&\label{eqFy}=\frac{\omega}{4}\text{Im}\left[\langle\delta m_{\pm,\phi}\delta m^*_{\pm,\theta}\rangle\right]\sin\phi,\\
	\mathscr{F}^{z}_{\pm}(\omega)&\label{eqFz}=\frac{\omega}{2} \text{Im}\left[\langle\delta m_{+,\phi}\delta m^*_{-,\phi}\rangle\right]_{\pm}\sin(2\phi).
\end{align}

The explicit expression for each contribution in Eq. (\ref{eq:scresult}) obeys
\begin{align}
\text{Im}\left[\langle\delta m_{+,\phi}\delta m^*_{+,\theta}\rangle\right]&=\frac{\omega B h^2_{+,\phi}}{\left|\det\left[\mathbb{M}_+\right]\right|^2},\\
\text{Im}\left[\langle\delta m_{-,\phi}\delta m^*_{-,\theta}\rangle\right]&=\frac{\omega Dh^2_{-,\phi}}{\left|\det\left[\mathbb{M}_-\right]\right|^2},\\
\text{Im}\left[\langle\delta m_{+,\phi}\delta m^*_{-,\phi}\rangle\right]_+&=\Delta\alpha\frac{\omega B\left(BD+\omega^2\right)\left(\omega^2_{-}-\omega^2\right)h^2_{+,\phi}}{\left|\det\left[\mathbb{M}_+\right]\right|^2\left|\det\left[\mathbb{M}_-\right]\right|^2},\\
\text{Im}\left[\langle\delta m_{+,\phi}\delta m^*_{-,\phi}\rangle\right]_-&=\Delta\alpha\frac{\omega D \left(BD+\omega^2\right)\left(\omega^2-\omega^2_+\right)h^2_{-,\phi}}{\left|\det\left[\mathbb{M}_+\right]\right|^2\left|\det\left[\mathbb{M}_-\right]\right|^2}.
\end{align}

\end{document}